\documentclass[11pt]{article}%
\usepackage{amsfonts}
\usepackage{amsmath}
\usepackage{amssymb}
\usepackage{graphicx}%
\begin{document}

\title{\hfill{\small UB-ECM-PF-04/07}\\\ \\Non-relativistic strings and branes\\as non-linear realizations of Galilei groups}
\author{Jan Brugues$^{\dagger}$, Thomas Curtright$^{\S }$, Joaquim Gomis$^{\dagger}$,
and Luca Mezincescu$^{\S }$
\and \ \\$^{\dagger}${\small Departament ECM, Facultat de F\'{\i}sica}\\{\small Institut de F\'{\i}sica d'Altes Energies and CER for Astrophysics}\\{\small Particle Physics and Cosmology, Universitat de Barcelona}\\{\small Diagonal 647, E-08028 Barcelona, Spain}\\\ \\$^{\S }${\small Department of Physics, University of Miami}\\{\small Coral Gables, Florida 33124-8046 USA}}
\maketitle

\begin{abstract}
We construct actions for non-relativistic strings and membranes purely as
Wess-Zumino terms of the underlying Galilei groups.

\end{abstract}

\section{Introduction}

Recently, a closed non-relativistic (NR) string with a non-trivial spectrum of
excitations was constructed \cite{GomisOoguri,DanielssonEtAl}. \ The
construction was motivated by non-commutative open string (NCOS) theories
\cite{SeibergEtAl,GopakumarEtAl} in $1+1$ dimensions \cite{KlebanovMaldacena}.
\ Here we would like to elucidate the symmetries and the geometrical structure
of the NR string. \ On the one hand, our goal will be to generalize the study
of the free non-relativistic particle action as a Wess-Zumino (WZ) term of the
ordinary Galilei group \cite{GauntlettEtAl}. \ On the other hand, our analysis
will parallel the study of the relativistic Green-Schwarz string action which
contains a WZ term \cite{HenneauxMezincescu} predicated on the non-trivial
third cohomology group of $\left(  N=2\ \mathrm{SuperPoincar\acute{e}}\right)
/SO(9,1)$ \cite{deAzcarragaEtAl1990}. \ After treating the string case, we
will then discuss the extension to non-relativistic d-branes.

\section{Non-relativistic strings}

The contraction of the Poincar\'{e} group associated with the NR string in $n$
spacetime dimensions is obtained by letting $c\rightarrow\infty$ after
rescaling the coordinates
\begin{equation}
x^{0}\rightarrow cx^{0}\equiv ct\ ,\ \ \ x^{1}\rightarrow cx^{1}\equiv
cx\ ,\ \ \ x^{a}\rightarrow X^{a}\;,a=\{2,...,n-1\}
\end{equation}
as well as rescaling the Poincar\'{e} generators%
\begin{align}
P^{0}  &  \rightarrow{H/c\ ,\ \ \ }P^{1}\rightarrow P{/c}\nonumber\\
J_{0a}  &  \rightarrow cJ_{0a}\equiv cK_{a}\ ,\ \ \ J_{1a}\rightarrow
cJ_{1a}\equiv cJ_{a}\nonumber\\
P_{a}  &  \rightarrow P_{a}\ ,\ \ \ J_{01}\rightarrow J_{01}\equiv
K\ ,\ \ \ J_{ab}\rightarrow M_{ab}%
\end{align}
The contracted algebra is then
\begin{align}
\lbrack M_{ab},M_{cd}]  &  =i\left(  \delta_{ac}M_{bd}+\delta_{bd}%
M_{ac}-\delta_{ad}M_{bc}-\delta_{bc}M_{ad}\right) \nonumber\\
\lbrack K_{a},M_{cd}]  &  =i\left(  \delta_{ad}K_{c}-\delta_{ac}K_{d}\right)
\;,\;\;\;[K_{a},K]=iJ_{a}\nonumber\\
\lbrack J_{a},M_{cd}]  &  =i\left(  \delta_{ad}J_{c}-\delta_{ac}J_{d}\right)
\;,\;\;\;[J_{a},K]=iK_{a}\nonumber\\
\lbrack M_{ab},P^{c}]  &  =i\left(  \delta_{ac}P_{b}-\delta_{bc}P_{a}\right)
\nonumber\\
\lbrack K,H]  &  =iP\;,\;\;\;[K,P]=iH\nonumber\\
\lbrack K_{a},H]  &  =iP_{a}\;,\;\;\;[J_{a},P]=iP_{a} \label{StringGalilei}%
\end{align}

Applying the general techniques \cite{IvanovOgievetsky} for non-linear
realizations of spacetime symmetries, as have been applied to
\emph{relativistic} membranes and supermembranes \cite{WestAndOthers}, we
consider the coset element
\begin{equation}
g=\exp\left(  -itH+ixP\right)  \exp\left(  iX^{a}P_{a}+iv^{b}K_{b}+i\theta
^{c}J_{c}\right)
\end{equation}
where $H,\ P$ are the unbroken translations, and $X^{a}\left(  t,x\right)
,\ v^{a}\left(  t,x\right)  ,\ \theta^{a}\left(  t,x\right)  $ are Goldstone
fields associated with the broken generators $P^{a},\ K^{a},\ J^{a}$,
respectively. The stability group is generated by the transverse rotations
$J^{ab},$ and by $K$. \ The transformation properties of $X^{a}\left(
t,x\right)  ,\ v^{a}\left(  t,x\right)  ,\ \theta^{a}\left(  t,x\right)  $ are
given (up to rotations) by
\begin{align}
t^{\prime}  &  =t_{0}+t\cosh v_{0}+x\sinh v_{0}\;,\;\;\;x^{\prime}%
=x_{0}+t\sinh v_{0}+x\cosh v_{0}\nonumber\\
v^{a}~^{\prime}  &  =v_{0}^{a}+v^{a}\cosh v_{0}+\theta^{a}\sinh v_{0}%
\ ,\ \ \ \theta^{a}~^{\prime}=\theta_{0}^{a}+v^{a}\sinh v_{0}+\theta^{a}\cosh
v_{0}\nonumber\\
X^{a}~^{\prime}  &  =X^{a}+X_{0}^{a}+\left(  t_{0}+t\cosh v_{0}+x\sinh
v_{0}\right)  v_{0}^{a}-\left(  x_{0}+t\sinh v_{0}+x\cosh v_{0}\right)
\theta_{0}^{a} \label{transformations}%
\end{align}
The Maurer-Cartan one-form $\Omega=-ig^{-1}dg$ is
\begin{align}
\Omega &  =K_{a}dv^{a}+J_{a}d\theta^{a}+P_{a}\left(  -v^{a}dt+\theta
^{a}dx+dX^{a}\right)  -Hdt+Pdx\nonumber\\
&  \equiv K_{a}\omega_{K_{a}}+J_{a}\omega_{J_{a}}+P_{a}\omega_{P_{a}}%
-H\omega_{H}+P\omega_{P}%
\end{align}
From this, we construct the closed, invariant three-form
\begin{equation}
\Omega_{3}=\omega_{K_{b}}\wedge\omega_{P_{b}}\wedge\omega_{P}-\omega_{J_{b}%
}\wedge\omega_{P_{b}}\wedge\omega_{H} \label{String3form}%
\end{equation}
Moreover, it is easy to obtain a two-form \textquotedblleft
potential\textquotedblright\ $\Phi_{2}$ such that $\Omega_{3}=d\Phi_{2}$,
namely%
\begin{equation}
\Phi_{2}={\frac{1}{{2}}}\left(  \theta^{2}-v^{2}\right)  dt\wedge
dx+v^{a}dX^{a}\wedge dx-\theta^{a}dX^{a}\wedge dt \label{String2form}%
\end{equation}
modulo addition of any closed or exact 2-form. $\ \Phi_{2}$ cannot be written
in terms of the left-invariant one forms, so it is not invariant under the
action of the group, and therefore the third cohomology group of the Galilei
group as given by (\ref{StringGalilei}) is not trivial.

There are two immediate developments. \ (I) \ We can construct an extended
algebra with new non-trivial commutation relations. \ In part, these would be
given by%
\begin{equation}
\lbrack K_{a},P_{b}]=i\delta_{ab}Z\ ,\ \ \ [K_{a},J_{b}]=i\delta
_{ab}W\ ,\ \ \ [P,K_{a}]=iN_{a}\ .
\end{equation}
Note that the presence of the \emph{central}\ \emph{charges} $Z\ $and $W$
break the $SO(1,1)$ invariance, since, for example,$\;[K_{a},P_{b}%
]=i\delta_{ab}Z\;$but$\;[J_{a},P_{b}]=0$. \ We will explore this further
elsewhere \cite{Brugues2appear}. \ 

(II) \ We can construct a WZ action given by $S_{WZ}=T\int\ast\Phi_{2}$ where
$\ast\Phi_{2}$ is the pullback of the two-form on the world sheet and
$T=1/\left(  2\pi l_{s}^{2}\right)  $ is the tension of the string. \ This WZ
action takes the explicit form
\begin{gather}
S_{WZ}=T\int\ast dt\wedge dx\left(  v^{a}\partial_{t}X^{a}+\theta^{a}%
\partial_{x}X^{a}+{\frac{1}{{2}}}\left(  \theta^{2}-v^{2}\right)  \right) \\
=T\int d\tau d\sigma\left(  \left(
\begin{array}
[c]{cc}%
v^{a} & \theta^{a}%
\end{array}
\right)  \left(
\begin{array}
[c]{cc}%
\partial_{\sigma}x & -\partial_{\tau}x\\
-\partial_{\sigma}t & \partial_{\tau}t
\end{array}
\right)  \left(
\begin{array}
[c]{c}%
\partial_{\tau}X^{a}\\
\partial_{\sigma}X^{a}%
\end{array}
\right)  +{\frac{1}{{2}}}\left(  \theta^{2}-v^{2}\right)  \left(
\partial_{\tau}t\partial_{\sigma}x-\partial_{\sigma}t\partial_{\tau}x\right)
\right) \nonumber
\end{gather}
If we fix the static gauge, $t=\tau$ and $x=\alpha\sigma$, $\alpha$ being a
parameter, and we eliminate the nondynamical Goldstone fields, we get
\cite{GarciaEtAl}
\begin{equation}
S_{WZ}=T\int dtdx\left(  \frac{1}{2}\left(  \partial_{t}X^{a}\right)
^{2}-\frac{1}{2}\left(  \partial_{x}X^{a}\right)  ^{2}\right)
\end{equation}
The transverse coordinates are a collection of free, massless fields on the world-sheet.

If we compute the Noether charges associated with the transformations
(\ref{transformations}), we find for the central and topological\ elements in
the gauge fixed form the formal expressions
\begin{equation}
Z=\int d\sigma\ \frac{\partial x}{\partial\sigma}\ ,\ \ \ W=\int
d\sigma\ \sigma\ \frac{\partial x}{\partial\sigma}\ ,\ \ \ N_{a}=\int
d\sigma\ \frac{\partial X_{a}}{\partial\sigma}\ .
\end{equation}
If $Z$ should be different from zero we need our space to be homologically
non-trivial, for example $S^{1}\times R^{n-1}$. \ For a closed string the
coordinate $x$ would then be wrapped \cite{CremmerScherk} around $S^{1}$,
which implies $x=2\pi Rk\sigma$, $\sigma\in\left[  0,1\right]  $, where $k$ is
the winding number and $R$ is the radius of $S^{1}$. \ Similarly, if we have a
homologically trivial transverse space $N_{a}$ is zero, but if we consider
some directions of the transverse space as tori, say, we will have some of the
$N_{a}\neq0$.

\section{Non-relativistic d-branes}

World-volume (longitudinal) dimensions are labeled by $x^{\alpha}$ with
$\alpha=0,1,\cdots,d$ and Lorentzian metric $\eta_{\alpha\beta}=diag\left(
-1,+1,\cdots,+1\right)  $, while the remaining (transverse) dimensions are
labeled by $X^{a}$ with $a=1,\cdots,D$ and Euclidean metric $\delta_{ab}$.

The group elements in the coset are%
\begin{equation}
g=\exp\left(  ix^{\alpha}p_{\alpha}\right)  \exp\left(  iX^{a}P_{a}+iv^{\alpha
a}K_{\alpha a}\right)
\end{equation}
As before, the algebra, without central extensions\footnote{We will include
central and topological charges \cite{deAzcarragaEtAl1989} in the algebra
later. \ See (\ref{CentroTopo}) below, and \cite{Brugues2appear}.}, is easily
abstracted from the contracted\ ($c\rightarrow\infty$) Galilean correspondence%
\begin{align}
p_{\alpha} &  \sim-i\frac{\partial}{\partial x^{\alpha}}\ ,\ \ \ P_{a}%
\sim-i\frac{\partial}{\partial X^{a}}\ ,\ \ \ K_{\alpha a}\sim-ix_{\alpha
}\frac{\partial}{\partial X^{a}}\nonumber\\
m_{\alpha\beta} &  \sim-i\left(  x_{\alpha}\frac{\partial}{\partial x^{\beta}%
}-x_{\beta}\frac{\partial}{\partial x^{\alpha}}\right)  \ ,\ \ \ M_{ab}%
\sim-i\left(  X_{a}\frac{\partial}{\partial X^{b}}-X_{b}\frac{\partial
}{\partial X^{a}}\right)
\end{align}
Thus the relevant non-vanishing commutators are%
\begin{equation}
\left[  K_{\alpha a},p_{\beta}\right]  =i\eta_{\alpha\beta}P_{a}%
\end{equation}
as well as $\left[  K_{\alpha a},m_{\beta\gamma}\right]  =i\eta_{\alpha\gamma
}K_{\beta a}-i\eta_{\alpha\beta}K_{\gamma a}\;,\;\;\left[  K_{\alpha a}%
,M_{bc}\right]  =i\delta_{ac}K_{\alpha b}-i\delta_{ab}K_{\alpha c}\ $, etc.
\ This leads to the one-form%
\begin{align}
\Omega &  =-ig^{-1}dg\nonumber\\
&  =p_{\alpha}\ dx^{\alpha}+P_{a}\left(  dX^{a}+v_{\alpha a}\ dx^{\alpha
}\right)  +K_{\alpha a}\ dv^{\alpha a}\nonumber\\
&  \equiv p_{\alpha}\omega^{\alpha}+P_{a}\omega^{a}+K_{\alpha a}\omega^{\alpha
a}%
\end{align}
with component one-forms defined as\footnote{The string case is obtained by
writing $v^{b}=v^{0b}$ and $\theta^{b}=v^{1b}$.}%
\begin{equation}
\omega^{\alpha}=dx^{\alpha}\ ,\ \ \ \omega^{a}=dX^{a}+v_{\alpha}%
^{a}\ dx^{\alpha}\ ,\ \ \ \omega^{\alpha a}=dv^{\alpha a}%
\end{equation}
The Maurer-Cartan equations involving the non-vanishing structure constants,
$f_{K_{\alpha a},p_{\beta},P_{b}}=\eta_{\alpha\beta}\delta_{ab}$, are given
here by $d\omega^{a}=\omega^{\alpha a}\wedge\omega_{\alpha}$.

We now construct from the component one-forms a closed, invariant $d+2$ form
\begin{equation}
\Omega_{d+2}=\frac{\left(  -1\right)  ^{d}}{d!}\ \varepsilon_{\alpha_{1}%
\cdots\alpha_{d}\beta}\ \omega^{\alpha_{1}}\wedge\cdots\wedge\omega
^{\alpha_{d}}\wedge\omega^{\beta b}\wedge\omega^{b}=d\Phi_{d+1}
\label{BraneForm}%
\end{equation}
with $\varepsilon_{01\cdots d}\equiv+1=-\varepsilon^{01\cdots d}$, to produce
a Wess-Zumino brane action%
\begin{equation}
\mathbb{A=}T_{d}\int_{\mathcal{M}_{d+2}}\ast\Omega_{d+2}=T_{d}\int
_{\mathcal{M}_{d+1}\ =\ \partial\mathcal{M}_{d+2}}\ast\Phi_{d+1},
\end{equation}
where $T_{d}$ is the tension of the d-brane. \ The boundary $\mathcal{M}%
_{d+1}=\partial\mathcal{M}_{d+2}$ is the world-volume of the evolving brane.
\ Straightforward calculation shows that an appropriate choice for the brane
form potential is%
\begin{equation}
\Phi_{d+1}=\frac{1}{d!}\ \varepsilon_{\alpha_{1}\cdots\alpha_{d}\beta
}\ dx^{\alpha_{1}}\wedge\cdots\wedge dx^{\alpha_{d}}\wedge\left(  v^{\beta
b}dX^{b}+\frac{1}{2d+2}\ v_{\gamma}^{b}v^{\gamma b}dx^{\beta}\right)
\label{BraneFormPotential}%
\end{equation}
Recall the $a,\ b,$ etc. indices are contracted with Euclidean metric, while
$\alpha,\ \beta,\ $etc. are handled with Lorentz metric.

As a special case, reconsider the NR string, with $d=1$. \ We have%
\begin{align}
\Omega_{3}  &  =\varepsilon_{\alpha\beta}\ \omega^{\alpha}\wedge\omega
^{b}\wedge\omega^{\beta b}=\varepsilon_{\alpha\beta}\ dx^{\alpha}\wedge\left(
dX^{b}+v_{\gamma}^{b}\ dx^{\gamma}\right)  \wedge dv^{\beta b}=d\Phi
_{2}\nonumber\\
\Phi_{2}  &  =\varepsilon_{\alpha\beta}\ dx^{\alpha}\wedge\left(  v^{\beta
b}dX^{b}+\frac{1}{4}\ v_{\gamma}^{b}v^{\gamma b}dx^{\beta}\right)
\end{align}
in agreement with (\ref{String3form}) and (\ref{String2form}). \ We note in
passing that $\Omega_{3}$ is an apparent modification of the usual torsion
3-form for non-linearly realized (compact) semi-simple Lie groups
\cite{BraatenCurtrightZachos} when expressed in terms of the component forms
(i.e. vielbeins)\footnote{Naively, this other 3-form, $\widetilde{\Omega}_{3}%
$, would be built from the component forms by using the structure constants
$f_{K_{\alpha a},p_{\beta},P_{b}}$ given earlier, and differs from $\Omega
_{3}$ by replacement of $\varepsilon_{\alpha\beta}$ with $\eta_{\alpha\beta}$.
\ For an arbitrary d-brane, it would be%
\[
\widetilde{\Omega}_{3}=\eta_{\alpha\beta}\delta_{ab}\ \omega^{\alpha}%
\wedge\omega^{a}\wedge\omega^{\beta b}=dx_{\alpha}\wedge\left(  dX^{b}%
+v_{\beta}^{b}\ dx^{\beta}\right)  \wedge dv^{\alpha b}%
\]
Remarkably, this 3-form is \emph{not} closed, for arbitrary $d$, but$\ $rather
gives $d\widetilde{\Omega}_{3}=\left(  dx_{\alpha}\wedge dv^{\alpha b}\right)
\wedge\left(  dx_{\beta}\wedge dv^{\beta b}\right)  $.}. \ 

If we fix the static gauge $x^{\alpha}=\sigma^{\alpha}$, the pullback to the
d-brane world-volume goes as follows.%
\begin{equation}
\left.  \Phi_{d+1}\right\vert _{X\left(  x\right)  }=\frac{1}{d!}%
\ \varepsilon_{\alpha_{1}\cdots\alpha_{d}\beta}\left(  -\varepsilon
^{\alpha_{1}\cdots\alpha_{d}\gamma}v^{\beta b}\frac{\partial X^{b}}{\partial
x^{\gamma}}-\frac{1}{2d+2}\ v_{\gamma}^{b}v^{\gamma b}\varepsilon^{\alpha
_{1}\cdots\alpha_{d}\beta}\right)  dx^{0}\wedge dx^{1}\wedge\cdots\wedge
dx^{d}%
\end{equation}
That is to say%
\begin{equation}
\left.  \Phi_{d+1}\right\vert _{X\left(  x\right)  }=\left(  v^{\gamma b}%
\frac{\partial X^{b}}{\partial x^{\gamma}}+\frac{1}{2}\ v^{\gamma b}v_{\gamma
}^{b}\right)  dx^{0}\wedge dx^{1}\wedge\cdots\wedge dx^{d}%
\end{equation}
So the action is
\begin{equation}
\mathbb{A=}T_{d}\int_{\mathcal{M}_{d+1}}\left(  v^{\gamma b}\frac{\partial
X^{b}}{\partial x^{\gamma}}+\frac{1}{2}\ v^{\gamma b}v_{\gamma}^{b}\right)
dx^{0}\wedge dx^{1}\wedge\cdots\wedge dx^{d}%
\end{equation}
Eliminating the auxiliary Goldstone fields $v^{\gamma b}$ using their
equations of motion, $v_{\gamma}^{b}=-\frac{\partial X^{b}}{\partial
x^{\gamma}}$, gives the gauge fixed\ action%
\begin{equation}
\mathbb{A=}T_{d}\int_{\mathcal{M}_{d+1}}\left(  -\frac{1}{2}\ \eta
^{\alpha\beta}\frac{\partial X^{a}}{\partial x^{\alpha}}\frac{\partial X^{a}%
}{\partial x^{\beta}}\right)  dx^{0}\wedge dx^{1}\wedge\cdots\wedge dx^{d}
\label{gfixed}%
\end{equation}
So, in a straightforward generalization of the string case given previously,
the $X^{a}$ considered as functions of the $x^{\alpha}$ are \emph{free
massless fields on the world-volume}.

In terms of another world-volume parameterization,$\ z^{\alpha}$,
$\alpha=0,1,\cdots,d$, and $z_{\beta}=\eta_{\beta\alpha}z^{\alpha}\ $, we have
$dx^{0}\wedge dx^{1}\wedge\cdots\wedge dx^{d}=\left\vert M\right\vert
\ dz^{0}\wedge\cdots\wedge dz^{d}$ where%
\begin{align}
M_{\beta}^{\ \alpha}  &  \equiv\frac{\partial}{\partial z^{\beta}}\ x^{\alpha
}\ ,\ \ \ M_{\beta\alpha}=\frac{\partial}{\partial z^{\beta}}\ x_{\alpha
}\ ,\ \ \ \frac{\partial X^{b}}{\partial x^{\gamma}}=\left(  M^{-1}\right)
_{\gamma\beta}\frac{\partial X^{b}}{\partial z_{\beta}}\nonumber\\
\left(  M^{-1}\right)  _{\alpha\beta}  &  =\frac{1}{\left\vert M\right\vert
}\ \text{$\mathfrak{C}$}\left(  M\right)  _{\beta\alpha}\ ,\ \ \ \left\vert
M\right\vert =\det M_{\beta}^{\ \alpha}=\varepsilon_{\alpha_{0}\alpha
_{1}\cdots\alpha_{d}}\frac{\partial x^{\alpha_{0}}}{\partial z^{0}}%
\frac{\partial x^{\alpha_{1}}}{\partial z^{1}}\cdots\frac{\partial
x^{\alpha_{d}}}{\partial z^{d}}%
\end{align}
with $\mathfrak{C}\left(  M\right)  $ the usual matrix of cofactors,
$\mathfrak{C}\left(  M\right)  _{\beta\alpha}=\left(  -1\right)
^{\alpha+\beta}\det\left(  \left.  M\right\vert _{\substack{\text{remove
}\beta\text{th row }\\\text{and }\alpha\text{th column }}}\right)  $.
\ Explicitly%
\begin{equation}
\text{$\mathfrak{C}$}\left(  M\right)  _{\beta\alpha}=\frac{-1}{d!}%
~\varepsilon_{\beta\beta_{1}\cdots\beta_{d}}~\varepsilon_{\alpha\alpha
_{1}\cdots\alpha_{d}}~\frac{\partial x^{\alpha_{1}}}{\partial z_{\beta_{1}}%
}~\frac{\partial x^{\alpha_{2}}}{\partial z_{\beta_{2}}}~\cdots~\frac{\partial
x^{\alpha_{d}}}{\partial z_{\beta_{d}}} \label{ExplicitCofactor}%
\end{equation}
In this arbitrary parameterization%
\begin{equation}
\Phi_{d+1}=\left(  v^{\gamma b}\left(  M^{-1}\right)  _{\gamma\beta}%
\frac{\partial X^{b}}{\partial z_{\beta}}+\frac{1}{2}\ v^{\gamma b}v_{\gamma
}^{b}\right)  \ \left\vert M\right\vert \ dz^{0}\wedge\cdots\wedge dz^{d}%
\end{equation}
The action in an arbitrary gauge is therefore%
\begin{equation}
\mathbb{A=}T_{d}\int_{\mathcal{M}_{d+1}}\left(  v^{\gamma b}%
\ \text{$\mathfrak{C}$}\left(  M\right)  _{\beta\gamma}\frac{\partial X^{b}%
}{\partial z_{\beta}}+\frac{1}{2}\ v^{\gamma b}v_{\gamma}^{b}\ \left\vert
M\right\vert \right)  \ dz^{0}\wedge\cdots\wedge dz^{d}%
\end{equation}
Eliminating once again the Goldstone auxiliaries using their equations of
motion, $v_{\gamma b}=-\left(  M^{-1}\right)  _{\gamma\alpha}\frac{\partial
X_{b}}{\partial z_{\alpha}}$, gives (for $d=2$, see \cite{GarciaEtAl})%
\begin{equation}
\mathbb{A=}T_{d}\int_{\mathcal{M}_{d+1}}\left(  \frac{-1}{2\left\vert
M\right\vert }\ \text{$\mathfrak{C}$}\left(  M\right)  _{\alpha\gamma}%
\ \eta^{\gamma\delta}\ \text{$\mathfrak{C}$}\left(  M\right)  _{\beta\delta
}\ \frac{\partial X^{b}}{\partial z_{\alpha}}\frac{\partial X^{b}}{\partial
z_{\beta}}\right)  \ dz^{0}\wedge\cdots\wedge dz^{d} \label{dAction}%
\end{equation}
The quadratic in cofactors can be written more explicitly using
(\ref{ExplicitCofactor}) above.%
\begin{equation}
\text{$\mathfrak{C}$}\left(  M\right)  _{\alpha\gamma}\ \eta^{\gamma\delta
}\ \text{$\mathfrak{C}$}\left(  M\right)  _{\beta\delta}=\frac{-1}%
{d!}~\varepsilon_{\alpha\alpha_{1}\cdots\alpha_{d}}~\varepsilon_{\beta
\beta_{1}\cdots\beta_{d}}~\frac{\partial x^{\gamma_{1}}}{\partial
z_{\alpha_{1}}}\frac{\partial x_{\gamma_{1}}}{\partial z_{\beta_{1}}}%
~\frac{\partial x^{\gamma_{2}}}{\partial z_{\alpha_{2}}}\frac{\partial
x_{\gamma_{2}}}{\partial z_{\beta_{2}}}~\cdots~\frac{\partial x^{\gamma_{d}}%
}{\partial z_{\alpha_{d}}}\frac{\partial x_{\gamma_{d}}}{\partial z_{\beta
_{d}}}%
\end{equation}
Rather deceptively, (\ref{dAction}) looks like the action for a nonlinear,
interacting theory coupling the transverse dimensions to the world-volume
variables. \ Of course, in the gauge $x^{\alpha}=z^{\alpha}$, we have
$M_{\alpha\beta}=\eta_{\alpha\beta}$, and we recover a free-field action as in
the string case -- a simplification typical of re-parameterization invariance.

The Noether charges for the d-brane, in the gauge fixed form (\ref{gfixed}),
are given at any world-volume time as
\begin{gather}
P_{a}=\int d^{d}x~\Pi_{a}\ ,\ \ \ K_{\alpha a}=\int d^{d}x~\left(  x_{\alpha
}\Pi_{a}-\eta_{\alpha0}~X_{a}\right) \nonumber\\
p_{0}={-}\int d^{d}x~\left(  \tfrac{1}{2}\Pi_{a}\Pi_{a}+\tfrac{1}{2}%
\partial_{i}X_{a}\partial_{i}X_{a}\right)  \equiv\int d^{d}x~\mathcal{P}%
_{0}\ ,\ \ \ p_{i}=-\int d^{d}x~\Pi_{a}\partial_{i}X_{a}\equiv\int
d^{d}x~\mathcal{P}_{i}\nonumber\\
m_{\alpha\beta}=\int d^{d}x~\left(  x_{\alpha}\mathcal{P}_{\beta}-x_{\beta
}\mathcal{P}_{\alpha}\right)  \ ,\ \ \ M_{ab}=\int d^{d}x~\left(  X_{a}\Pi
_{b}-X_{b}\Pi_{a}\right)
\end{gather}
where $a,b=1,\cdots,D$, and $\alpha,\beta=0,1,\cdots,d$, but $i=1,\cdots,d$.
\ Here, $\Pi_{a}$ is the canonically conjugate momentum associated with
$X_{a}$, so $\left.  \left[  X_{a}\left(  x\right)  ,\Pi_{b}\left(  x^{\prime
}\right)  \right]  \right\vert _{x_{0}=x_{0}^{\prime}}=i\delta_{ab}%
\delta^{\left(  d\right)  }\left(  x-x^{\prime}\right)  $. \ 

For the general d-brane, the commutation relations are just a straightforward
generalization of the non-relativistic string case. \ However, a greater
variety of central and topological charges may appear in the d-brane algebra,
for example as
\begin{equation}
\lbrack P_{a},K_{\alpha b}]=i\eta_{0\alpha}\delta_{ab}Z\ ,\ \ \ [K_{0a}%
,K_{ib}]=i\delta_{ab}W_{i}\ ,\ \ \ [p_{i},K_{0a}]=iN_{ia}\ .
\label{CentroTopo}%
\end{equation}
These central and topological charges are given formally for the d-brane
by\footnote{In the last relation in (\ref{CentroTopo}) and/or
(\ref{FormalCentroTopo}), we have \emph{assumed} there are no conjugate
momentum winding numbers, so we have set\ $\int d^{d}x~\partial_{i}\Pi
_{a}\left(  x\right)  =0$. \ If otherwise, we would write $N_{ia}=\int\left(
\partial_{i}X_{a}+x_{0}\partial_{i}\Pi_{a}\right)  ~d^{d}x\ .$}
\begin{equation}
Z=\int d^{d}x\ ,\ \ \ W_{i}=\int x_{i}~d^{d}x\ ,\ \ \ N_{ia}=\int
\frac{\partial X_{a}}{\partial x^{i}}~d^{d}x\ . \label{FormalCentroTopo}%
\end{equation}
To avoid ambiguities in these formal expressions, we have checked the results
against various Jacobi identities. \ A full discussion of all Jacobi
identities is deferred \cite{Brugues2appear}.

\section{Conclusion}

We have shown how to construct non-relativistic string and brane actions from
the structure of the underlying Galilei group. \ We will investigate quantum
properties of these models and discuss the roles of their topological charges
in a subsequent paper \cite{Brugues2appear}. \ 

In certain cases, perhaps all, passive advections of non-relativistic strings
and membranes also result from a Wess-Zumino action when expressed in the
formalism of Nambu mechanics \cite{Curtright}. \ It would be interesting to
combine fully that other formalism with the group theoretic approach of the
present paper.

\paragraph{Acknowledgements:}

We acknowledge discussions with David Fairlie, Jaume Gomis, Kiyoshi Kamimura,
and Paul Townsend. \ This research was supported in part by NSF Awards 9870101
and 0303550, by European Commission RTN project HPNR-CT-2000-00131, by MCYT
FPA 2001-3598, and by CIRIT 2001SGR-00065. \ JB is supported by a grant from
Ministerio de Ciencia y Tecnologia. \ \ JG thanks the Physics Department,
University of Miami, for its hospitality while part of this work was done.

\end{document}